\documentclass[aps,10pt,pra,twocolumn,amsmath,amssymb,superscriptaddress,groupedaddress]{revtex4-1}

\usepackage[english]{babel} 
\usepackage[utf8]{inputenc}
\usepackage[T1]{fontenc}
\usepackage{graphicx}
\usepackage{epsfig}
\usepackage{graphicx}
\usepackage{subfigure}
\usepackage{todonotes}

\usepackage{hyperref}
\hypersetup{
	colorlinks=true,
	linkcolor=blue,			
	citecolor=green,		
	filecolor=magenta,		
	urlcolor=magenta		
}

\newcommand{\abs}[1]{\left\vert #1 \right\vert}
\newcommand{\erw}[1]{\left\langle #1 \right\rangle}
\newcommand{\komm}[2]{\left[ #1 , #2 \right]}

\renewcommand{\d}{\mathrm{d}}
\newcommand{\comma}{~,}
\newcommand{\fullstop}{~.}
\newcommand{\Tr}{\mathrm{Tr}}
\renewcommand{\Re}{\mathrm{Re}}
\newcommand{\kB}{\mathrm{k}_\mathrm{B}}

\begin{document}

\selectlanguage{english}

\def\ValueFigureIIOmegaMech{1.0}
\def\ValueFigureIIDelta{0.36}
\def\ValueFigureIIg0{0.36}
\def\ValueFigureIIalphaLaser{0.3}
\def\ValueFigureIIkappa{1.5}
\def\ValueFigureIIGammaMech{0.00125}
\def\ValueFigureIInPhonon{0.0}
\def\ValueFigureIINumberOfTrajectories{30}
\def\ValueFigureIIQuantile{70}
\def\ValueFigureIIBss{3.2}
\def\ValueFigureIIangle{0.6}

\def\ValueFigureIIIPseudoRealisticOmegaMech{1.0}
\def\ValueFigureIIIPseudoRealisticDelta{0.05}
\def\ValueFigureIIIPseudoRealisticg{0.16}
\def\ValueFigureIIIPseudoRealisticalphaLaser{0.1}
\def\ValueFigureIIIPseudoRealistickappa{0.1}
\def\ValueFigureIIIPseudoRealisticGammaMech{0.0006}
\def\ValueFigureIIIPseudoRealisticnPhonon{0}
\def\ValueFigureIIIPseudoRealisticOptimalAngle{0.15}
\def\ValueFigureIIIPseudoRealisticErwN{10.8}
\def\ValueFigureIIIPseudoRealisticScB{3.2}
\def\ValueFigureIIIPseudoRealisticScReBeta{0.14}

\def\ValueFigureIIIZeroMechanicalDampingOmegaMech{1.0}
\def\ValueFigureIIIZeroMechanicalDampingg{0.16}
\def\ValueFigureIIIZeroMechanicalDampingalphaLaser{0.35}
\def\ValueFigureIIIZeroMechanicalDampingGammaMech{0}
\def\ValueFigureIIIZeroMechanicalDampingnPhonon{0}

\def\ValueFigureIIIZeroMechanicalDampingDeltaA{0.05}
\def\ValueFigureIIIZeroMechanicalDampingkappaA{0.1}
\def\ValueFigureIIIZeroMechanicalDampingOptimalAngleA{0.0}
\def\ValueFigureIIIZeroMechanicalDampingErwNA{55.5}
\def\ValueFigureIIIZeroMechanicalDampingScBA{7.5}
\def\ValueFigureIIIZeroMechanicalDampingScReBetaA{0.47}

\def\ValueFigureIIIZeroMechanicalDampingDeltaB{0.21}
\def\ValueFigureIIIZeroMechanicalDampingkappaB{0.5}
\def\ValueFigureIIIZeroMechanicalDampingOptimalAngleB{0.0}
\def\ValueFigureIIIZeroMechanicalDampingErwNB{57.7}
\def\ValueFigureIIIZeroMechanicalDampingScBB{7.6}
\def\ValueFigureIIIZeroMechanicalDampingScReBetaB{0.15}

\def\ValueFigureIIIZeroMechanicalDampingDeltaC{0.37}
\def\ValueFigureIIIZeroMechanicalDampingkappaC{1.0}
\def\ValueFigureIIIZeroMechanicalDampingOptimalAngleC{0.9}
\def\ValueFigureIIIZeroMechanicalDampingErwNC{64.7}
\def\ValueFigureIIIZeroMechanicalDampingScBC{8.0}
\def\ValueFigureIIIZeroMechanicalDampingScReBetaC{0.05}

\def\ValueFigureIIINonZeroMechanicalDampingOmegaMech{1.0}
\def\ValueFigureIIINonZeroMechanicalDampingg{0.36}
\def\ValueFigureIIINonZeroMechanicalDampingalphaLaser{0.3}
\def\ValueFigureIIINonZeroMechanicalDampingGammaMech{0.00125}
\def\ValueFigureIIINonZeroMechanicalDampingnPhonon{0}

\def\ValueFigureIIINonZeroMechanicalDampingDeltaA{0.00}
\def\ValueFigureIIINonZeroMechanicalDampingkappaA{0.1}
\def\ValueFigureIIINonZeroMechanicalDampingOptimalAngleA{0.0}
\def\ValueFigureIIINonZeroMechanicalDampingErwNA{9.4}
\def\ValueFigureIIINonZeroMechanicalDampingScBA{3.2}
\def\ValueFigureIIINonZeroMechanicalDampingScReBetaA{0.39}

\def\ValueFigureIIINonZeroMechanicalDampingDeltaB{0.15}
\def\ValueFigureIIINonZeroMechanicalDampingkappaB{0.5}
\def\ValueFigureIIINonZeroMechanicalDampingOptimalAngleB{0.9}
\def\ValueFigureIIINonZeroMechanicalDampingErwNB{10.3}
\def\ValueFigureIIINonZeroMechanicalDampingScBB{3.2}
\def\ValueFigureIIINonZeroMechanicalDampingScReBetaB{0.21}

\def\ValueFigureIIINonZeroMechanicalDampingDeltaC{0.27}
\def\ValueFigureIIINonZeroMechanicalDampingkappaC{1.0}
\def\ValueFigureIIINonZeroMechanicalDampingOptimalAngleC{0.8}
\def\ValueFigureIIINonZeroMechanicalDampingErwNC{11.3}
\def\ValueFigureIIINonZeroMechanicalDampingScBC{3.2}
\def\ValueFigureIIINonZeroMechanicalDampingScReBetaC{0.09}

\def\ValueFigureIIINonZeroMechanicalDampingDeltaD{0.36}
\def\ValueFigureIIINonZeroMechanicalDampingkappaD{1.5}
\def\ValueFigureIIINonZeroMechanicalDampingOptimalAngleD{0.6}
\def\ValueFigureIIINonZeroMechanicalDampingErwND{11.8}
\def\ValueFigureIIINonZeroMechanicalDampingScBD{3.2}
\def\ValueFigureIIINonZeroMechanicalDampingScReBetaD{0.05}

\def\ValueFigureIIINonZeroMechanicalDampingDeltaE{1.44}
\def\ValueFigureIIINonZeroMechanicalDampingkappaE{2.0}
\def\ValueFigureIIINonZeroMechanicalDampingOptimalAngleE{0.4}
\def\ValueFigureIIINonZeroMechanicalDampingErwNE{22.2}
\def\ValueFigureIIINonZeroMechanicalDampingScBE{4.6}
\def\ValueFigureIIINonZeroMechanicalDampingScReBetaE{0.01}

\def\ValueFigureIIINonZeroMechanicalDampingDeltaF{1.22}
\def\ValueFigureIIINonZeroMechanicalDampingkappaF{2.5}
\def\ValueFigureIIINonZeroMechanicalDampingOptimalAngleF{0.3}
\def\ValueFigureIIINonZeroMechanicalDampingErwNF{15.3}
\def\ValueFigureIIINonZeroMechanicalDampingScBF{3.8}
\def\ValueFigureIIINonZeroMechanicalDampingScReBetaF{0.01}

\def\ValueFigureIIINonZeroMechanicalDampingDeltaG{0.74}
\def\ValueFigureIIINonZeroMechanicalDampingkappaG{3.0}
\def\ValueFigureIIINonZeroMechanicalDampingOptimalAngleG{0.3}
\def\ValueFigureIIINonZeroMechanicalDampingErwNG{8.0}
\def\ValueFigureIIINonZeroMechanicalDampingScBG{2.6}
\def\ValueFigureIIINonZeroMechanicalDampingScReBetaG{0.01}

\def\ValueFigureIIIEfficiencyOmegaMech{1.0}
\def\ValueFigureIIIEfficiencyDelta{0.36}
\def\ValueFigureIIIEfficiencyg{0.36}
\def\ValueFigureIIIEfficiencyalphaLaser{0.3}
\def\ValueFigureIIIEfficiencykappa{1.5}
\def\ValueFigureIIIEfficiencyGammaMech{0.00125}
\def\ValueFigureIIIEfficiencynPhonon{0}
\def\ValueFigureIIIEfficiencyOptimalAngle{0.6}
\def\ValueFigureIIIEfficiencyErwN{11.8}
\def\ValueFigureIIIEfficiencyScB{3.2}
\def\ValueFigureIIIEfficiencyScReBeta{0.05}

\def\ValueFigureIIIQuantile{70}

\def\ValueFigureIVOmegaMech{1.0}
\def\ValueFigureIVkappa{1.5}
\def\ValueFigureIVGammaMech{0.005}
\def\ValueFigureIVnPhonon{0}
\def\ValueFigureIVgAlphaLaser{0.108}

\def\ValueFigureIVDeltaA{0.07}
\def\ValueFigureIVgA{0.72}
\def\ValueFigureIValphaLaserA{0.15}
\def\ValueFigureIVOptimalAngleA{0.6}
\def\ValueFigureIVErwNA{2.0}
\def\ValueFigureIVScBA{0.3}
\def\ValueFigureIVScReBetaA{0.03}

\def\ValueFigureIVDeltaB{0.44}
\def\ValueFigureIVgB{0.36}
\def\ValueFigureIValphaLaserB{0.30}
\def\ValueFigureIVOptimalAngleB{0.2}
\def\ValueFigureIVErwNB{7.0}
\def\ValueFigureIVScBB{2.5}
\def\ValueFigureIVScReBetaB{0.04}

\def\ValueFigureIVDeltaC{0.55}
\def\ValueFigureIVgC{0.24}
\def\ValueFigureIValphaLaserC{0.45}
\def\ValueFigureIVOptimalAngleC{0.3}
\def\ValueFigureIVErwNC{15.6}
\def\ValueFigureIVScBC{4.0}
\def\ValueFigureIVScReBetaC{0.05}

\def\ValueFigureIVDeltaD{0.58}
\def\ValueFigureIVgD{0.18}
\def\ValueFigureIValphaLaserD{0.60}
\def\ValueFigureIVOptimalAngleD{0.3}
\def\ValueFigureIVErwND{28.8}
\def\ValueFigureIVScBD{5.4}
\def\ValueFigureIVScReBetaD{0.07}

\def\ValueFigureIVDeltaE{0.61}
\def\ValueFigureIVgE{0.144}
\def\ValueFigureIValphaLaserE{0.75}
\def\ValueFigureIVOptimalAngleE{0.4}
\def\ValueFigureIVErwNE{43.8}
\def\ValueFigureIVScBE{6.8}
\def\ValueFigureIVScReBetaE{0.08}

\def\ValueFigureIVQuantile{70}

\def\ValueFigureVOmegaMech{1.0}
\def\ValueFigureVkappa{1.5}
\def\ValueFigureVg{0.36}
\def\ValueFigureVGammaMech{0.00125}
\def\ValueFigureValphaLaser{0.3}

\def\ValueFigureVnPhononA{0.0}
\def\ValueFigureVDeltaA{0.36}
\def\ValueFigureVOptimalAngleA{0.6}
\def\ValueFigureVErwNA{11.8}
\def\ValueFigureVScBA{3.2}
\def\ValueFigureVScReBetaA{0.05}

\def\ValueFigureVnPhononB{0.01}
\def\ValueFigureVDeltaB{0.36}
\def\ValueFigureVOptimalAngleB{0.6}
\def\ValueFigureVErwNB{11.8}
\def\ValueFigureVScBB{3.2}
\def\ValueFigureVScReBetaB{0.05}

\def\ValueFigureVnPhononC{0.05}
\def\ValueFigureVDeltaC{0.36}
\def\ValueFigureVOptimalAngleC{0.6}
\def\ValueFigureVErwNC{11.8}
\def\ValueFigureVScBC{3.2}
\def\ValueFigureVScReBetaC{0.05}

\def\ValueFigureVnPhononD{0.1}
\def\ValueFigureVDeltaD{0.36}
\def\ValueFigureVOptimalAngleD{0.6}
\def\ValueFigureVErwND{11.8}
\def\ValueFigureVScBD{3.2}
\def\ValueFigureVScReBetaD{0.05}

\def\ValueFigureVnPhononE{0.25}
\def\ValueFigureVDeltaE{0.35}
\def\ValueFigureVOptimalAngleE{0.6}
\def\ValueFigureVErwNE{11.8}
\def\ValueFigureVScBE{3.2}
\def\ValueFigureVScReBetaE{0.05}

\def\ValueFigureVnPhononF{0.5}
\def\ValueFigureVDeltaF{0.34}
\def\ValueFigureVOptimalAngleF{0.6}
\def\ValueFigureVErwNF{11.9}
\def\ValueFigureVScBF{3.2}
\def\ValueFigureVScReBetaF{0.05}

\def\ValueFigureVnPhononG{1.0}
\def\ValueFigureVDeltaG{0.34}
\def\ValueFigureVOptimalAngleG{0.6}
\def\ValueFigureVErwNG{12.1}
\def\ValueFigureVScBG{3.2}
\def\ValueFigureVScReBetaG{0.05}

\def\ValueFigureVnPhononH{2.0}
\def\ValueFigureVDeltaH{0.33}
\def\ValueFigureVOptimalAngleH{0.6}
\def\ValueFigureVErwNH{12.7}
\def\ValueFigureVScBH{3.2}
\def\ValueFigureVScReBetaH{0.05}

\def\ValueFigureVQuantile{70}

\title{Unraveling nonclassicality in the optomechanical instability}

\author{Martin Koppenh\"ofer}
\affiliation{Department of Physics, University of Basel, Klingelbergstrasse 82, CH-4056 Basel, Switzerland}

\author{Christoph Bruder}
\affiliation{Department of Physics, University of Basel, Klingelbergstrasse 82, CH-4056 Basel, Switzerland}

\author{Niels L\"orch}
\affiliation{Department of Physics, University of Basel, Klingelbergstrasse 82, CH-4056 Basel, Switzerland}

\date{\today}

\begin{abstract}
	Conditional dynamics due to continuous optical measurements has successfully been applied for state reconstruction and feedback cooling in optomechanical systems. 
	In this article, we show that the same measurement techniques can be used to unravel nonclassicality in optomechanical limit cycles. 
	In contrast to unconditional dynamics, our approach gives rise to nonclassical limit cycles even in the sideband-unresolved regime, where the cavity decay rate exceeds the mechanical frequency.
	We predict a significant reduction of the mechanical amplitude fluctuations for realistic experimental parameters. 
\end{abstract}

\maketitle

\section{Introduction}
In recent years, optomechanical experiments have started to enter the quantum regime. 
Sideband cooling of the mechanical subsystem to the quantum ground state \cite{nature.475.359,nature.478.89}, sensing of mechanical motion at the standard quantum limit \cite{NatureNano.4.820,science.339.801,science.344.1486}, quantum state transfer between the optical and mechanical subsystems \cite{nature.482.63,nature.495.210}, and phonon lasing \cite{PhysRevLett.104.083901,nphys.5.909,nature.520.522} have been demonstrated experimentally. 
Theoretical studies \cite{PhysRevLett.104.053601,PhysRevLett.109.253601,PhysRevA.88.053828,PhysRevA.94.063835} have led to the prediction \cite{PhysRevX.4.011015} that the phonon distribution of such an optomechanical phonon laser can be nonclassical if the system is operated in the resolved-sideband regime.
However, an experimental observation of this feature is still missing.

Continuous measurements, such as homodyne detection or photon counting, can provide information on the state of a system \cite{bk-WisemanMilburn-QMC}. 
These measurements give rise to a conditional time evolution since the state of the system at any given time depends on the previous measurement results. 
Real-time state reconstruction using these measurement results has been experimentally demonstrated both in the regime of negligible optical backaction  
and in the quantum regime \cite{EPJD.22.131,PhysRevLett.111.163602,PhysRevLett.114.223601}. 
In a second step, the obtained knowledge of the system state can be used to implement feedback mechanisms to, e.g., cool the motion of the system \cite{PhysRevLett.80.688,PhysRevA.60.2700,PhysRevB.68.235328,Nature.524.325}. 
Squeezing of the mechanical motion of a levitated nanosphere in the presence of sideband-cooling and Markovian feedback has been studied theoretically \cite{NewJPhys.17.073019}.

In this article we consider the opposite limit, i.e., an optomechanical system driven into mechanical limit-cycle motion by a blue-detuned laser. 
We show that a continuous measurement on the optical cavity can be used to reveal nonclassicality of the mechanical state by reducing its mechanical amplitude fluctuations. 
We quantify this reduction using the mechanical Fano factor, which is the mean-square fluctuation of the phonon number normalized to the phonon-number expectation value. 

In contrast to existing proposals based on unconditional dynamics \cite{PhysRevLett.104.053601,PhysRevX.4.011015}, which require to be operated in the sideband-resolved regime in order to obtain nonclassical mechanical states, our approach opens the exciting possibility of nonclassical self-oscillations in the sideband-unresolved regime.
This comes at the cost of obtaining a time-dependent, stochastically fluctuating Fano factor because of the conditional system dynamics. 
We characterize the magnitude of these fluctuations and show that the conditional Fano factor can become smaller than unity, notably even in the sideband-unresolved regime, where it has been proven that the Fano factor observed for unconditional dynamics is always larger than unity \cite{PhysRevX.4.011015}.

This article is structured as follows: 
In Sec.~\ref{sec:Methods} we introduce our system and the different parameter regimes that we investigate. 
Numerical results are shown in Sec.~\ref{sec:Results} and discussed in Sec.~\ref{sec:Discussion}. 
Finally, we conclude in Sec.~\ref{sec:Conclusion}.

\section{Methods}
\label{sec:Methods}
\begin{figure}
	\includegraphics[width=.46\textwidth]{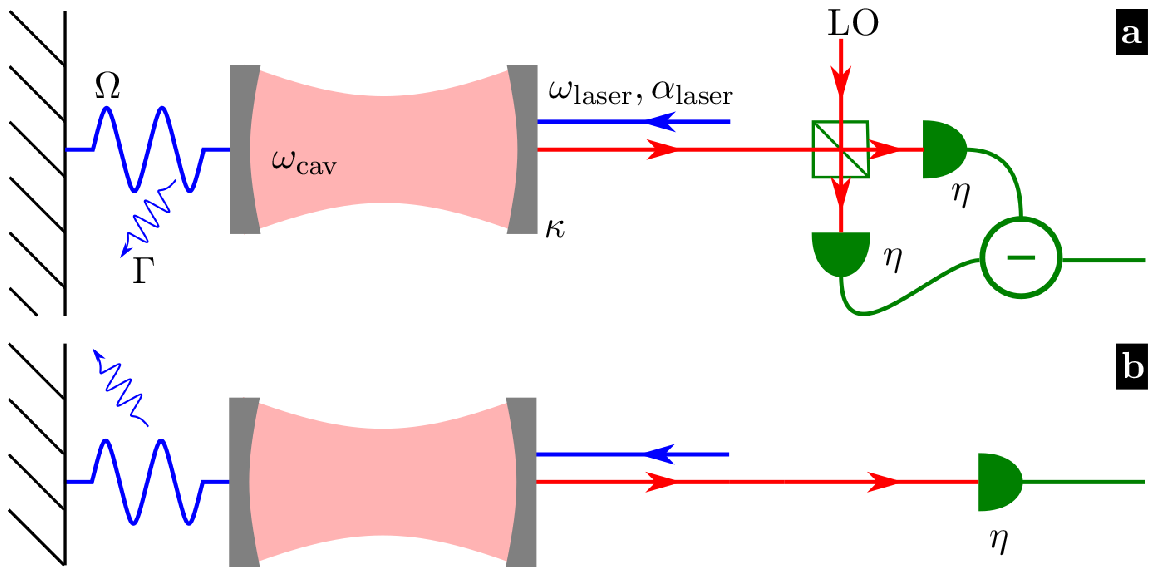}
	\includegraphics[width=.46\textwidth]{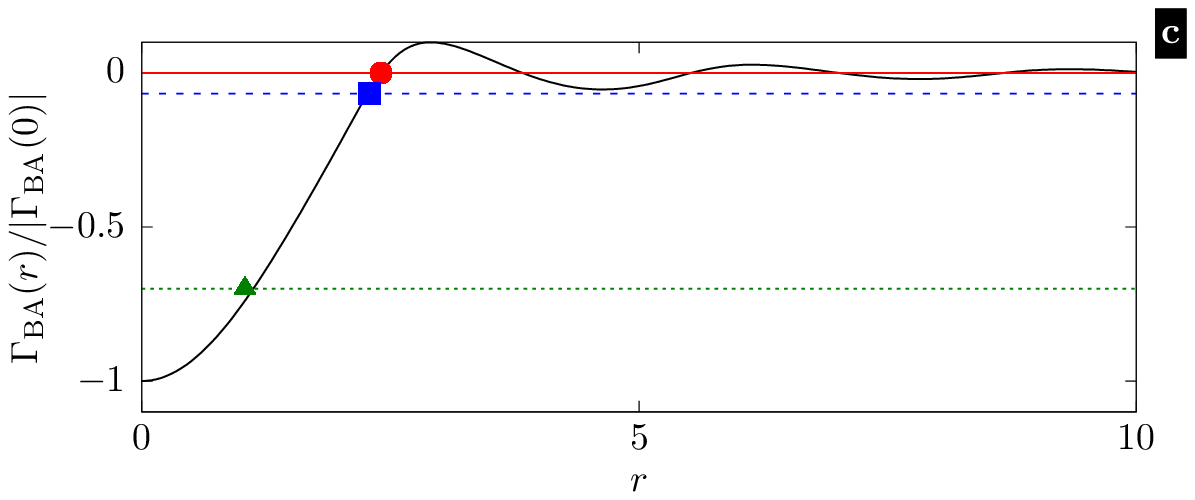}
	\caption{
		Sketch of the considered setup. 
		An optomechanical system is monitored by a continuous measurement, either homodyne detection (a) or photon counting (b). 
		A blue-detuned laser drive is applied to the optomechanical system to induce mechanical limit cycles at an amplitude $B_\mathrm{ss}$.  
		(c) Optically induced damping $\Gamma_\mathrm{BA}$ as a function of $r = 2 g_0 B_\mathrm{ss}/\Omega$ in the regime $\abs{\Delta} < \Omega$.
		Here, $\Delta$ is the detuning between the frequencies of the laser drive and the cavity, $\Omega$ is the frequency of the mechanical oscillator, and $g_0$ is the bare single-photon optomechanical coupling strength. 
		Limit-cycle positions for the different parameter sets of the mechanical damping considered in this paper (cf.\ main text) are indicated by solid markers. 
		The colors and marker symbols correspond to the ones of the data sets in Fig.~\ref{fig:3}
	}
	\label{fig:1}
\end{figure}

We consider an optomechanical system described by the quantum master equation
\begin{align}
	\dot{\rho} = 
		&- i \komm{H}{\rho} 
		+ \kappa \mathcal{D}[a] \rho \nonumber\\
		&+ (n_\mathrm{ph} + 1) \Gamma \mathcal{D}[b] \rho
		+ n_\mathrm{ph} \Gamma \mathcal{D}[b^\dagger] \rho \comma
		\label{eqn:System:QuantumMasterEquation} \\
	H = 
		&- \Delta a^\dagger a 
		+ \Omega b^\dagger b \nonumber \\
		&- g_0 a^\dagger a \left( b^\dagger + b \right) 
		+ \alpha_\mathrm{laser} \left( a^\dagger + a \right) \comma 
		\label{eqn:System:Hamiltonian} 
\end{align}
where $a$ and $b$ are the annihilation operators of an optical photon and a mechanical phonon, respectively, and $\Delta = \omega_\mathrm{laser} - \omega_\mathrm{cav}$ is the detuning of the frequency $\omega_\mathrm{laser}$ of the laser drive with respect to the cavity frequency $\omega_\mathrm{cav}$. 
The Lindblad dissipators are defined by $\mathcal{D}[O] \rho = O \rho O^\dagger - \frac{1}{2} O^\dagger O \rho - \frac{1}{2} \rho O^\dagger O$.
The mechanical and optical damping rates and the thermal mechanical phonon number are denoted by $\Gamma$, $\kappa$, and $n_\mathrm{ph}$, respectively, and we put $\hbar=1$. 
The symbols $\Omega$, $g_0$, and  $\alpha_\mathrm{laser}$ denote the mechanical resonance frequency, the bare optomechanical coupling strength, and the amplitude of the laser drive, respectively. 
A continuous measurement, i.e., photon counting or a homodyne detection of the optical quadrature $a e^{i \varphi} + a^\dagger e^{- i \varphi}$, is performed on the output port of the optical cavity. We will focus on the case where the angle $\varphi$ is chosen to minimize the Fano factor.
The photodetectors used there are assumed to have a detection efficiency $\eta$. 
A sketch of the setup introducing the parameters is shown in Figs.~\ref{fig:1}(a) and (b). 

In this article, we assume a blue-detuned laser drive to be applied to the optomechanical system, $\Delta > 0$, to drive the system into mechanical limit-cycle motion. 
This motion is of the form \cite{PhysRevLett.104.053601,PhysRevLett.96.103901} 
\begin{align}
	\erw{b} = \overline{\beta}_\mathrm{ss} + B_\mathrm{ss} e^{-i (\Omega t + \phi)} \comma
	\label{eqn:System:MechanicalMotion}
\end{align}
where $B_\mathrm{ss}$ is the steady-state amplitude of the limit-cycle motion and $\overline{\beta}_\mathrm{ss}$ is a constant offset. 
The value $B_\mathrm{ss}$ is obtained by equating the negative mechanical damping rate $- \Gamma$ with the optically induced damping rate $\Gamma_\mathrm{BA}(r)$, defined as \cite{PhysRevLett.104.053601,PhysRevLett.96.103901} 
\begin{align}
	\Gamma_\mathrm{BA}(r) &= 2 \kappa g_0^2 \abs{\alpha_\mathrm{laser}}^2 \sum_{n = - \infty}^\infty \frac{J_n(r) J_{n+1}(r)}{r h_n h_{n+1}} \comma
\end{align}
where $J_n$ is the $n$-th Bessel function of the first kind and the dimensionless amplitude $r$ is defined by $r = 2 g_0 B_\mathrm{ss}/\Omega$.
This damping rate is sketched in Fig.~\ref{fig:1}(c). 
We also introduced the abbreviation $h_n = \kappa^2/4 + (n \Omega + \Delta + 2 g_0 \Re [ \overline{\beta}_\mathrm{ss} ])^2$. 
Amplitude fluctuations around the steady-state value, $\delta B(t) = B(t) - B_\mathrm{ss}$, decay exponentially, $\delta B(t) = \delta B(0) e^{- \Gamma_\mathrm{rel} t/2}$, where the relaxation rate is given by (cf.\ Ref.~\onlinecite{PhysRevLett.104.053601})
\begin{align}
	\Gamma_\mathrm{rel} 
	&= 2 \kappa g_0^2 \abs{\alpha_\mathrm{laser}}^2 \sum_{n = -\infty}^\infty \frac{J_n'(r) J_{n+1}(r) + J_n(r) J_{n+1}'(r)}{h_n h_{n+1}} \nonumber \\ 
	&+ \Gamma \fullstop
	\label{eqn:System:AmplitudeRelaxationRate}
\end{align}

We consider three different parameter regimes of the mechanical damping $\Gamma$, schematically represented in Fig.~\ref{fig:1}(c) by horizontal lines:

(i) Negligible mechanical damping (solid red line). 
In this case, the theory developed in Ref.~\onlinecite{PhysRevX.4.011015} is directly applicable. 
Furthermore, a measurement at perfect detection efficiency, $\eta = 1$, drives the system into a pure state. 
As $\Gamma_\mathrm{BA}(r)$ oscillates around zero, limit cycles are possible at each positively-sloped root of $\Gamma_\mathrm{BA}(r)$. 
Quantum fluctuations and noise induced by the optical decay may drive the system across the regions of positive damping separating these limit-cycle positions. 
The value of the maximum damping rate in these regions decreases if $\kappa$ is increased, therefore, for larger $\kappa$ the system can show multistability between several limit cycles or it may eventually become unstable and does not feature a limit cycle at all. 
In effect, all parameters considered in this paper are in the sideband-resolved regime if the mechanical damping rate is set to zero. 

(ii) A nonzero mechanical damping that is small compared to $\abs{\Gamma_\mathrm{BA}(0)}$ at mechanical amplitude zero, but large enough such that there is a unique limit cycle (dashed blue line).
In this regime, there is no restriction on the values of $\kappa$. 

The two parameter regimes introduced above are advantageous for numerical simulations, since, for an appropriate choice of $g_0$, we can avoid multistability by widely separating the limit cycle amplitudes in Fock space while the fluctuations in Fock number $n$ are still sufficiently small to allow us to restrict the Hilbert space dimension to a numerically feasible size.  
However, so far optomechanical limit cycle experiments operate at very small values of $g_0/\kappa$,  the ratio $\kappa/\Omega$ can be smaller than unity, but the cooperativities are not much larger than unity, such that their limit cycle is situated in the parabolic region of the curve of $\Gamma_\mathrm{BA}(r)$ \cite{nature.520.522}.

(iii) Therefore, we also investigate a mechanical damping of the order of $\abs{\Gamma_\mathrm{BA}(0)}$ (dotted green line) to provide a result in a corresponding parameter regime. 
However, we increase the value of $g_0$ as compared to the experiment because the Hilbert space dimension scales inversely proportional to $g_0$. 

To quantify the nonclassicality of the mechanical limit cycle, we calculate the mechanical Fano factor
\begin{align}
	F = \frac{\erw{n^2} - \erw{n}^2}{\erw{n}} \comma
	\label{eqn:System:MechanicalFanoFactor}
\end{align}
where $n = b^\dagger b$ denotes the phonon number operator. 
It is a measure of the mean-square amplitude fluctuations normalized to the amplitude expectation value. 
A coherent (classical) state corresponds to a Fano factor of unity, whereas thermal states or states broadened by other noise feature a Fano factor larger than unity. A Fano factor less than unity indicates a nonclassical, sub-Poissonian squeezed mechanical state. 

Applying homodyne detection on the optical cavity output is a continuous measurement of the optical quadrature $a e^{i \varphi} + a^\dagger e^{-i \varphi}$. 
Likewise, a photon counting measurement continuously monitors single-photon losses of the optical cavity. 
These measurements introduce a conditional time evolution, since the state of the system depends on all previous measurement results. 
To simulate a continuous measurement, the unconditional quantum master equation~\eqref{eqn:System:QuantumMasterEquation} is turned into a stochastic quantum master equation for the density operator $\rho$, which describes random quantum trajectories of the system and can be solved numerically \cite{bk-WisemanMilburn-QMC}. 
In the case of photon counting, the system dynamics consists of a deterministic time evolution that is interrupted by occasional quantum jumps at random times. 
In the case of homodyne detection, the system dynamics is a white noise process so that a random measurement result has to be taken into account in every time step. 
The eigenvalues of the deterministic part of the differential equation have a large imaginary component, such that explicit integration algorithms are unstable even for relatively small timesteps. 
Therefore, we use a semi-implicit Milstein algorithm, which is implicit with respect to the deterministic part of the time evolution \cite{bk-KloedenPlaten}, to integrate the stochastic differential equation in the case of homodyne detection. 
This algorithm has already been implemented in the QuTiP package \cite{cpc.183.1760}, which we use for all numerical calculations in this article.
To solve the stochastic differential equation for photon counting, we have implemented a fourth-order implicit Runge-Kutta algorithm \cite{bk-Rannacher}.

The steady-state system dynamics can be recovered from the stochastic calculation by an average over many quantum trajectories. 
Hence, the averages of observables, e.g., the averaged phonon number expectation value $\overline{\erw{n}}$, do not change their value in the presence of continuous measurements. 
However, knowledge acquired by the continuous measurement drives the density matrix on each individual quantum trajectory closer to a pure state, i.e., the von Neumann entropy $S = - \kB \Tr \left( \rho \ln \rho \right)$ decreases.
Likewise, the values of functions of observables, e.g., the Fano factor, may take values different from their steady-state values. 
In the following, we numerically study how both homodyne detection and photon counting reduce the mechanical Fano factor of the system. 
We analyze the impact of inefficient detection, $\eta < 1$, of finite mechanical temperature $n_\mathrm{ph} > 0$, and of the optomechanical coupling strength $g_0$.

\section{Results}
\label{sec:Results}
\begin{figure}
	\includegraphics[width=.42\textwidth]{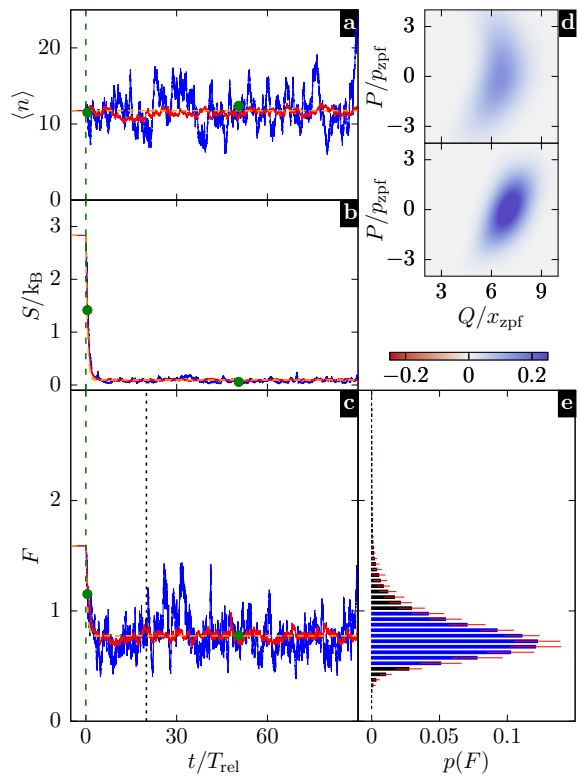}
	\caption{
          Evolution of (a) the phonon-number expectation value $\erw{n}$, (b) the von Neumann entropy $S$, and (c) the Fano factor $F$ under homodyne detection for a single trajectory (solid blue [black in print] curves).
          The corresponding averages over \ValueFigureIINumberOfTrajectories~trajectories are shown as solid red [dark gray] curves, and the theoretical expectations in the limit of an average over infinitely many trajectories, Eqs.~\eqref{eqn:Results:St} and~\eqref{eqn:Results:Ft}, respectively, are shown as dashed orange [light gray] curves. 
		Time is given in units of the relaxation time $T_\mathrm{rel} = 1/\Gamma_\mathrm{rel}$.
		At times $t < 0$ the system is assumed to be in its steady state, homodyne detection is switched on at $t=0$. 
		Subfigure (d) shows the Wigner function of the mechanical state at two different times indicated by green dots on the trajectories in a frame rotating at $\Omega$, the earlier time being shown at the top.
		The zero-point fluctuations are defined by $x_\mathrm{zpf} = 1/\sqrt{2 m \Omega}$ and $p_\mathrm{zpf} = \sqrt{m \Omega/2}$, where $m$ denotes the mass of the movable mirror. 
		(e) Distribution of the Fano factor for times larger than the data-acquisition start time indicated by the dotted black line in (c).
		The data includes all \ValueFigureIINumberOfTrajectories~ trajectories. 
		The blue [dark gray] bars of the histogram comprise at least $\ValueFigureIIQuantile\,\%$ of the total probability.
		In the sideband-resolved regime, this region is asymmetrically distributed around the mean value. 
		Parameters are: $\Delta/\Omega = \ValueFigureIIDelta$, $g_0/\Omega = \ValueFigureIIg0$, $\kappa/\Omega = \ValueFigureIIkappa$,  $\Gamma/\Omega = \ValueFigureIIGammaMech$, $\alpha_\mathrm{laser}/\Omega = \ValueFigureIIalphaLaser$, $n_\mathrm{ph} = \ValueFigureIInPhonon$, $\varphi/\pi=\ValueFigureIIangle$, 
		and $\eta = 1$. 
		The value of $B_\mathrm{ss}$ characterizing the semiclassical solution is $B_\mathrm{ss} = \ValueFigureIIBss$.
	}
	\label{fig:2}
\end{figure}

Figures~\ref{fig:2}(a) to (c) show the phonon-number expectation value $\erw{n}$, the von Neumann entropy $S = - \kB \Tr (\rho \ln \rho)$, and the Fano factor $F$ for a quantum trajectory obtained by homodyne detection (solid blue [black in print] curves), as well as their mean values $\overline{\erw{n}}$, $\overline{S}$, and $\overline{F}$ obtained by averaging over many trajectories (solid red [dark gray] curves).
Time is given in units of the inverse amplitude relaxation rate, $T_\mathrm{rel} = 1/\Gamma_\mathrm{rel}$, cf.\ Eq.~\eqref{eqn:System:AmplitudeRelaxationRate}. 
The system is initialized in its steady state at times $t < 0$. 
At $t=0$, homodyne detection is switched on and $\erw{n}$, $S$, and $F$ evolve stochastically.
As expected, homodyne detection does not change the mean phonon-number expectation value $\overline{\erw{n}}$, but our increasing knowledge of the system state causes the mean von Neumann entropy $\overline{S}$ and the mean Fano factor $\overline{F}$ to be decreased.
Empirically, an exponential decay towards new conditional mean values is found, with a rate that is approximately twice the amplitude relaxation rate~\eqref{eqn:System:AmplitudeRelaxationRate}, 
\begin{align}
	\overline{S}(t) \approx (S_\mathrm{ss} - S_\mathrm{cond}) e^{- 2 \Gamma_\mathrm{rel} t} + S_\mathrm{cond} \comma 
	\label{eqn:Results:St}\\
	\overline{F}(t) \approx (F_\mathrm{ss} - F_\mathrm{cond}) e^{- 2 \Gamma_\mathrm{rel} t} + F_\mathrm{cond} \fullstop
	\label{eqn:Results:Ft}
\end{align}
This exponential decay is shown by dashed orange [light gray] curves. 
The value of $S_\mathrm{cond}$ depends on the strength of the remaining dissipative channels of the system. 
For zero mechanical damping and perfect detection, there is no unmonitored dissipation channel left and the system evolves into a pure entangled state, having zero von Neumann entropy. 
For nonzero mechanical damping or imperfect detection efficiency, there is an additional unmonitored decay channel such that the system evolves into a mixed state, having nonzero von Neumann entropy, $S_\mathrm{cond} \gtrsim 0$. 

In the case of homodyne detection, the value of $F_\mathrm{cond}$ depends on the measured quadrature, i.e., it is a function of the homodyne angle $\varphi$. 
In the following, all homodyne detection data is given at the optimal angle $\varphi$ that minimizes the value of $F_\mathrm{cond}$. 
Note that $\varphi$ and $\varphi + \pi$ effectively measure the same quadrature. 

In the limit $t \gg T_\mathrm{rel}$, the instantaneous Fano factor $F(t)$ of a quantum trajectory fluctuates around a constant mean value $F_\mathrm{cond}$. 
To quantify these fluctuations, we calculate the histogram of the Fano factor $F(t)$ over many trajectories for times $t$ larger than a data-acquisition start time indicated in Fig.~\ref{fig:2}(c) by a black dashed line. 
This histogram is shown in Fig.~\ref{fig:2}(e). 
In the following, we quantify its properties by three numbers:
(i) the mean Fano factor $F_\mathrm{cond}$, (ii) the probability $p(F < F_\mathrm{ss})$ to obtain a Fano factor smaller than  the steady-state value $F_\mathrm{ss}$, and (iii) the range of values of the Fano factor that contains at least $\ValueFigureIIQuantile\,\%$ of the total probability. 
This range is indicated by blue (dark gray in print) bars in the histogram and can be asymmetrically distributed around the mean value $F_\mathrm{cond}$ in the sideband-resolved regime. 
To determine this range, we calculate the cumulative distribution function of the histogram and exclude all bins that have a value smaller than $15\,\%$ or larger than $85\,\%$. 

\begin{figure}
	\includegraphics[width=.42\textwidth]{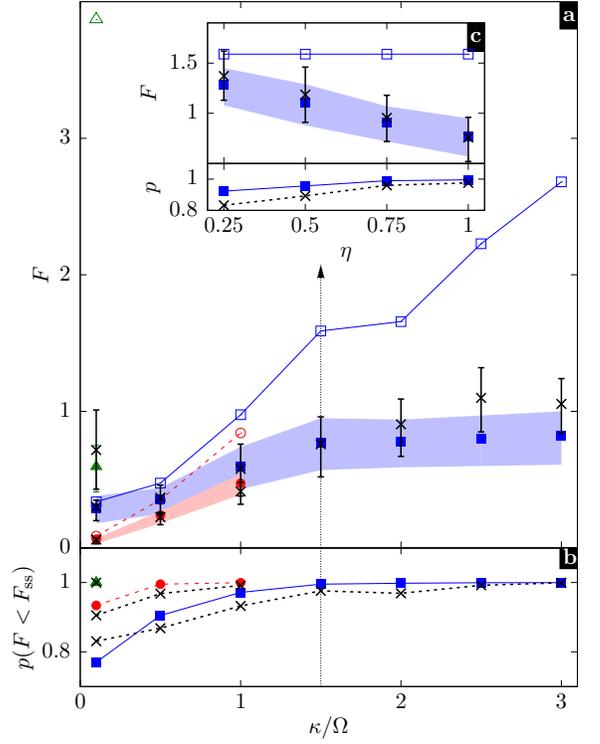}
	\caption{
		(a) Mean conditional Fano factor $F_\mathrm{cond}$ under a continuous measurement (solid markers), steady-state Fano factor $F_\mathrm{ss}$ (open markers) and (b) probability $p(F < F_\mathrm{ss})$ to observe a Fano factor smaller than $F_\mathrm{ss}$ as a function of the optical damping rate $\kappa$. 
		Red circles and dashed lines correspond to homodyne detection and zero mechanical damping. 
		Blue squares and solid lines correspond to homodyne detection and a mechanical damping small compared to $\abs{\Gamma_\mathrm{BA}(0)}$.
		Green triangles represent the case of homodyne detection and a mechanical damping rate large compared to $\abs{\Gamma_\mathrm{BA}(0)}$. 
		The data drawn with black crosses adjacent to the results for homodyne detection shows the corresponding results for photon counting. 
		The shaded regions and error bars represent the ranges of Fano factors that contain at least $\ValueFigureIIIQuantile\,\%$ of all counts. 
		All curves are taken at perfect detection efficiency, $\eta = 1$. 
		The detuning $\Delta$ and the homodyne detection angle $\varphi$ are chosen such that the steady-state Fano factor $F_\mathrm{ss}$ and its mean conditional value $F_\mathrm{cond}$ are minimal, respectively.
		The colors and marker symbols corresponding to the different parameter regimes are the same as the ones used in Fig.~\ref{fig:1}(c). 	
		The properties of the histogram in Fig.~\ref{fig:2}(e) are shown by the blue squares at $\kappa/\Omega = \ValueFigureIIkappa$. 
		(c) Influence of imperfect detection, $\eta < 1$, for $\Gamma/\Omega = \ValueFigureIIIEfficiencyGammaMech$ and $\kappa/\Omega = \ValueFigureIIIEfficiencykappa$. 
		Parameters are:
		$\kappa/\Omega = (\ValueFigureIIINonZeroMechanicalDampingkappaA, \ValueFigureIIINonZeroMechanicalDampingkappaB, \ValueFigureIIINonZeroMechanicalDampingkappaC, \ValueFigureIIINonZeroMechanicalDampingkappaD, \ValueFigureIIINonZeroMechanicalDampingkappaE, \ValueFigureIIINonZeroMechanicalDampingkappaF, \ValueFigureIIINonZeroMechanicalDampingkappaG)$ 
		and $n_\mathrm{ph} = \ValueFigureIIIZeroMechanicalDampingnPhonon$.
			red circles:
			$g_0/\Omega = \ValueFigureIIIZeroMechanicalDampingg$, 
			$\Gamma/\Omega = \ValueFigureIIIZeroMechanicalDampingGammaMech$, 
			$\alpha_\mathrm{laser}/\Omega = \ValueFigureIIIZeroMechanicalDampingalphaLaser$, 
			$\Delta/\Omega = (\ValueFigureIIIZeroMechanicalDampingDeltaA, \ValueFigureIIIZeroMechanicalDampingDeltaB, \ValueFigureIIIZeroMechanicalDampingDeltaC)$, 
			and $\varphi/\pi = (\ValueFigureIIIZeroMechanicalDampingOptimalAngleA, \ValueFigureIIIZeroMechanicalDampingOptimalAngleB, \ValueFigureIIIZeroMechanicalDampingOptimalAngleC)$;
			blue squares and black crosses:
			$g_0/\Omega = \ValueFigureIIINonZeroMechanicalDampingg$, 
			$\Gamma/\Omega = \ValueFigureIIINonZeroMechanicalDampingGammaMech$, 
			$\alpha_\mathrm{laser}/\Omega = \ValueFigureIIINonZeroMechanicalDampingalphaLaser$, 
			$\Delta/\Omega = (\ValueFigureIIINonZeroMechanicalDampingDeltaA, \ValueFigureIIINonZeroMechanicalDampingDeltaB, \ValueFigureIIINonZeroMechanicalDampingDeltaC, \ValueFigureIIINonZeroMechanicalDampingDeltaD, \ValueFigureIIINonZeroMechanicalDampingDeltaE, \ValueFigureIIINonZeroMechanicalDampingDeltaF, \ValueFigureIIINonZeroMechanicalDampingDeltaG)$, 
			and $\varphi/\pi = (\ValueFigureIIINonZeroMechanicalDampingOptimalAngleA, \ValueFigureIIINonZeroMechanicalDampingOptimalAngleB, \ValueFigureIIINonZeroMechanicalDampingOptimalAngleC, \ValueFigureIIINonZeroMechanicalDampingOptimalAngleD, \ValueFigureIIINonZeroMechanicalDampingOptimalAngleE, \ValueFigureIIINonZeroMechanicalDampingOptimalAngleF, \ValueFigureIIINonZeroMechanicalDampingOptimalAngleG)$; 
			green triangles:
			$g_0/\Omega = \ValueFigureIIIPseudoRealisticg$, 
			$\Gamma/\Omega = \ValueFigureIIIPseudoRealisticGammaMech$, 
			$\alpha_\mathrm{laser}/\Omega = \ValueFigureIIIPseudoRealisticalphaLaser$,  
			$\Delta/\Omega = \ValueFigureIIIPseudoRealisticDelta$, 
			and $\varphi/\pi = \ValueFigureIIIPseudoRealisticOptimalAngle$. 
		The corresponding values of $\erw{n}_\mathrm{ss}$ and $B_\mathrm{ss}$ are:
			red circles:
			$\erw{n}_\mathrm{ss} = (\ValueFigureIIIZeroMechanicalDampingErwNA, \ValueFigureIIIZeroMechanicalDampingErwNB, \ValueFigureIIIZeroMechanicalDampingErwNC)$ 
			and $B_\mathrm{ss} = (\ValueFigureIIIZeroMechanicalDampingScBA, \ValueFigureIIIZeroMechanicalDampingScBB, \ValueFigureIIIZeroMechanicalDampingScBC)$;
			blue circles:
			$\erw{n}_\mathrm{ss} = (\ValueFigureIIINonZeroMechanicalDampingErwNA, \ValueFigureIIINonZeroMechanicalDampingErwNB, \ValueFigureIIINonZeroMechanicalDampingErwNC, \ValueFigureIIINonZeroMechanicalDampingErwND, \ValueFigureIIINonZeroMechanicalDampingErwNE, \ValueFigureIIINonZeroMechanicalDampingErwNF, \ValueFigureIIINonZeroMechanicalDampingErwNG)$ 
			and $B_\mathrm{ss} = (\ValueFigureIIINonZeroMechanicalDampingScBA, \ValueFigureIIINonZeroMechanicalDampingScBB, \ValueFigureIIINonZeroMechanicalDampingScBC, \ValueFigureIIINonZeroMechanicalDampingScBD, \ValueFigureIIINonZeroMechanicalDampingScBE, \ValueFigureIIINonZeroMechanicalDampingScBF, \ValueFigureIIINonZeroMechanicalDampingScBG)$;
			green triangles:
			$\erw{n}_\mathrm{ss} = \ValueFigureIIIPseudoRealisticErwN$ and $B_\mathrm{ss} = \ValueFigureIIIPseudoRealisticScB$. 
		}
	\label{fig:3}
\end{figure}

Using these three figures of merit, we investigate the efficiency of the reduction of the Fano factor by continuous measurements compared to its steady-state value, for different optical damping rates $\kappa$, different types of continuous measurements, and damping rates in the three different regimes introduced in Sec.~\ref{sec:Methods}. 
Figures~\ref{fig:3}(a) and (b) summarize the results as a function of the optical decay rate $\kappa$, ranging from the sideband-resolved to the sideband-unresolved regime. 
Solid red circles represent $F_\mathrm{cond}$ and $p(F < F_\mathrm{ss})$, respectively, for homodyne detection and zero mechanical damping, $\Gamma/\Omega = 0$. 
The shaded region indicates the range of Fano factors that contains $\ValueFigureIIIQuantile\,\%$ of all counts. 
Open red circles connected by a dashed line represent the corresponding values of $F_\mathrm{ss}$.
For the given parameters there is no limit cycle in the region $\kappa \gtrsim \Omega$. 
Likewise, the curves indicated by blue squares and connected by solid lines represent the corresponding results for homodyne detection and a nonzero mechanical damping which is small compared to $\abs{\Gamma_\mathrm{BA}(0)}$. 
Finally, the green triangles show the results for homodyne detection and a mechanical damping of the order of $\abs{\Gamma_\mathrm{BA}(0)}$. 
For these three different parameter regimes, we also give the results obtained by photon counting, which are indicated by the black crosses with error bars or black crosses connected by dotted lines adjacent to the corresponding results for homodyne measurements.

As predicted in Refs.~\onlinecite{PhysRevLett.104.053601,PhysRevX.4.011015}, the steady-state Fano factor is smaller than unity in the sideband-resolved regime at zero mechanical damping. 
This prediction still holds for small mechanical damping, but is not applicable for a mechanical damping of the order of $\abs{\Gamma_\mathrm{BA}(0)}$. 
If $\kappa$ is increased towards the sideband-unresolved regime, the steady-state Fano factor grows and takes values much larger than unity. 
Homodyne detection or photon counting measurements decrease the conditional mean Fano factor $F_\mathrm{cond}$ with respect to $F_\mathrm{ss}$ for all considered values of $\Gamma$ and $\kappa$. 
Whereas $F_\mathrm{cond}$ grows in the sideband-resolved regime with increasing $\kappa$, it saturates to a value of about unity in the sideband-unresolved regime. 
In the sideband-resolved regime, $F_\mathrm{cond}$ depends only weakly on the homodyne angle $\varphi$ and homodyne detection and photon counting yield the same results within the statistical errors. 
Towards the sideband-unresolved regime or for large mechanical damping, however, the choice of an optimal homodyne angle $\varphi$ allows us to reach smaller values of $F_\mathrm{cond}$ than for photon counting.
The optimal homodyne angle is the one for which the measured optical quadrature oscillates at the highest frequency and is closest to a harmonic oscillation.

The probability $p(F < F_\mathrm{ss})$ to observe Fano factors smaller than the steady-state Fano factor under a continuous measurement increases towards the sideband-unresolved regime and approaches unity. 

The inset Fig.~\ref{fig:3}(c) shows the influence of the detection efficiency $\eta$ on $F_\mathrm{ss}$, $F_\mathrm{cond}$, and $p(F < F_\mathrm{ss})$ for $\kappa/\Omega = \ValueFigureIIIEfficiencykappa$. 
The smaller the detection efficiency the less information can be gained out of the continuous measurement. 
Therefore, $F_\mathrm{cond}$ tends towards the steady-state value $F_\mathrm{ss}$ for low detection efficiency.

\begin{figure}
	\includegraphics[width=.46\textwidth]{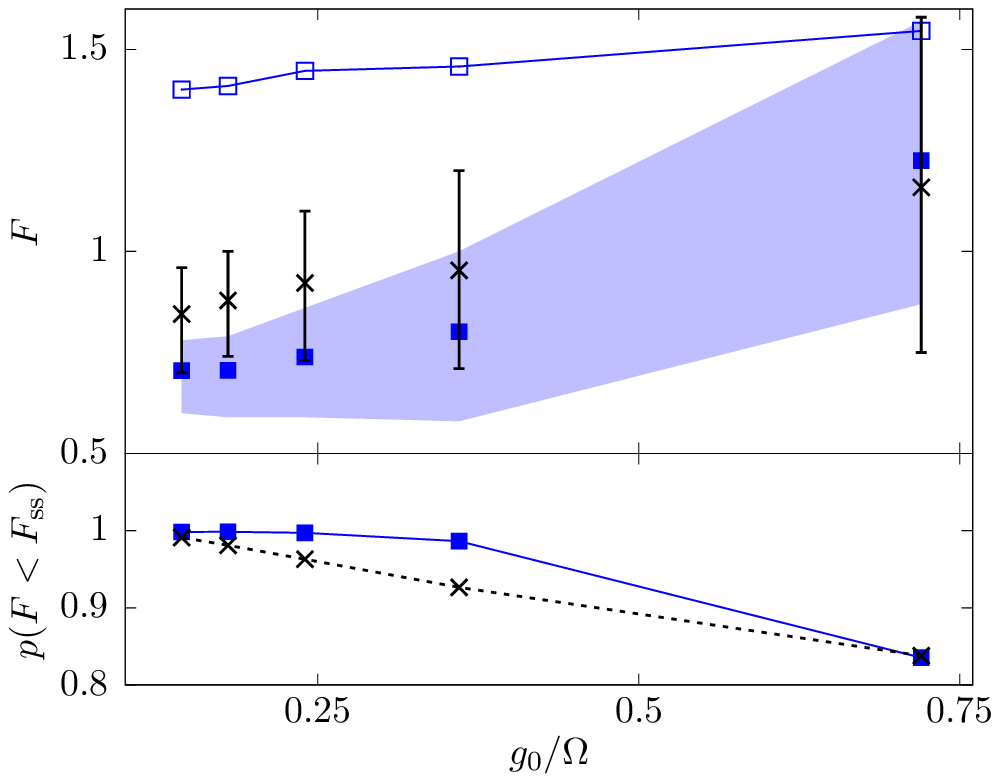}
	\caption{
		Influence of the optomechanical single-photon coupling strength $g_0$ on $F_\mathrm{ss}$, $F_\mathrm{cond}$, and $p(F < F_\mathrm{ss})$. 
		The product $g_0 \abs{\alpha_\mathrm{laser}} = \ValueFigureIVgAlphaLaser \,\Omega^2$ is kept constant. 
		The detuning $\Delta$ and the homodyne angle $\varphi$ are chosen such that the steady-state Fano factor $F_\mathrm{ss}$ and its mean conditional value $F_\mathrm{cond}$ are minimal, respectively.
		Filled blue squares indicate the results for homodyne detection, black crosses the ones for photon counting.
		The steady-state Fano factor $F_\mathrm{ss}$ is shown in open blue squares.
		Parameters are 
		$\kappa/\Omega = \ValueFigureIVkappa$, $\Gamma/\Omega = \ValueFigureIVGammaMech$, 
		$g_0/\Omega = (\ValueFigureIVgE, \ValueFigureIVgD, \ValueFigureIVgC, \ValueFigureIVgB, \ValueFigureIVgA)$, 
		$\alpha_\mathrm{laser}/\Omega = (\ValueFigureIValphaLaserE, \ValueFigureIValphaLaserD, \ValueFigureIValphaLaserC, \ValueFigureIValphaLaserB, \ValueFigureIValphaLaserA)$, 
		$\Delta/\Omega = (\ValueFigureIVDeltaE, \ValueFigureIVDeltaD, \ValueFigureIVDeltaC, \ValueFigureIVDeltaB, \ValueFigureIVDeltaA)$, 
		$n_\mathrm{ph} = \ValueFigureIVnPhonon$, and $\varphi/\pi = (\ValueFigureIVOptimalAngleE, \ValueFigureIVOptimalAngleD, \ValueFigureIVOptimalAngleC, \ValueFigureIVOptimalAngleB, \ValueFigureIVOptimalAngleA)$.
		The corresponding values of $\erw{n}_\mathrm{ss}$ and $B_\mathrm{ss}$ are
		$\erw{n}_\mathrm{ss} = (\ValueFigureIVErwNE, \ValueFigureIVErwND, \ValueFigureIVErwNC, \ValueFigureIVErwNB, \ValueFigureIVErwNA)$ and
		$B_\mathrm{ss} = (\ValueFigureIVScBE, \ValueFigureIVScBD, \ValueFigureIVScBC, \ValueFigureIVScBB, \ValueFigureIVScBA)$.
	}
	\label{fig:4}
\end{figure}
In Fig.~\ref{fig:4} we investigate the influence of the optomechanical single-photon coupling strength $g_0$ on the reduction of the Fano factor. 
To obtain comparable results, we rescale both $\abs{\alpha_\mathrm{laser}}$ and $g_0$ at a time such that their product $g_0 \abs{\alpha_\mathrm{laser}}$ is kept constant, because, in the limit of a small ratio of $g_0/\Omega$, the steady-state Fano factor is expected to be only a function of $g_0 \abs{\alpha_\mathrm{laser}}$ \cite{PhysRevX.4.011015}. 
Figure~\ref{fig:4} displays $F_\mathrm{ss}$, $F_\mathrm{cond}$, and $p(F < F_\mathrm{ss})$ as a function of the coupling strength $g_0$ and confirms this prediction. 
The mean Fano factor $F_\mathrm{cond}$ increases if the optomechanical coupling gets larger and approaches the steady-state value $F_\mathrm{ss}$. 
Likewise, the probability $p(F < F_\mathrm{ss})$ decreases. 

\begin{figure}
	\includegraphics[width=.46\textwidth]{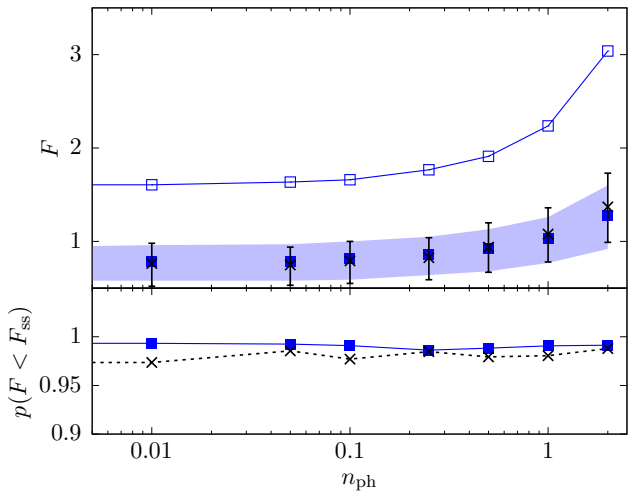}
	\caption{
		Influence of the mechanical temperature, expressed in terms of the thermal phonon number $n_\mathrm{ph}$, on $F_\mathrm{ss}$, $F_\mathrm{cond}$, and $p(F < F_\mathrm{ss})$. 
		The detuning $\Delta$ and the homodyne angle $\varphi$ are chosen such that the steady-state Fano factor $F_\mathrm{ss}$ and its mean conditional value $F_\mathrm{cond}$ are minimal, respectively.
		Filled blue squares indicate the results for homodyne detection, black crosses the ones for photon counting.
		The steady-state Fano factor $F_\mathrm{ss}$ is shown in open blue squares.
		Parameters are 
		$\kappa/\Omega = \ValueFigureVkappa$, 
		$\Gamma/\Omega = \ValueFigureVGammaMech$, 
		$g_0/\Omega = \ValueFigureVg$, 
		$\alpha_\mathrm{laser}/\Omega = \ValueFigureValphaLaser$, 
		$\Delta/\Omega = (\ValueFigureVDeltaB,\ValueFigureVDeltaC,\ValueFigureVDeltaD,\ValueFigureVDeltaE,\ValueFigureVDeltaF,\ValueFigureVDeltaG, \ValueFigureVDeltaH)$, 
		and $\varphi/\pi = \ValueFigureVOptimalAngleA$.
		The corresponding values of $\erw{n}_\mathrm{ss}$ and $B_\mathrm{ss}$ are
		$\erw{n}_\mathrm{ss} = 12$ and
		$B_\mathrm{ss} = \ValueFigureVScBA$.
	}
	\label{fig:5}
\end{figure}
Finally, Fig.~\ref{fig:5} shows the influence of the mechanical temperature, expressed in terms of the thermal phonon number $n_\mathrm{ph}$, on the reduction of the Fano factor. 
A reduction of $F_\mathrm{cond}$ compard to $F_\mathrm{ss}$ is observed for all considered temperatures. 
However, in order to observe a nonclassical Fano factor $F_\mathrm{cond} < 1$, a small effective mechanical thermal occupation $n_\mathrm{ph} \lesssim 1$ is required.

\section{Discussion}
\label{sec:Discussion}
The results shown in Fig.~\ref{fig:3} indicate that a continuous measurement on the cavity output of an optomechanical system decreases the mean mechanical Fano factor $F_\mathrm{cond}$ compared to the steady-state value $F_\mathrm{ss}$ in the absence of a continuous measurement.

The difference $F_\mathrm{ss} - F_\mathrm{cond}$ is particularly large in the sideband-unresolved regime and for a mechanical damping close to $\abs{\Gamma_\mathrm{BA}(0)}$. 
For parameters similar to the ones realized in current experiments (cf. green triangles in Fig.~\ref{fig:3}), the large steady-state Fano factor $F_\mathrm{ss} > 3$ is strongly reduced to $F_\mathrm{cond} < 1$, and we observe a nonclassical state with probability one. 
A similar, but less pronounced reduction is observed in the sideband-resolved regime. 
In the sideband-unresolved regime, we are able to reduce the Fano
factor from a steady-state value much larger than unity to a mean value $F_\mathrm{cond}$ close to unity. 
Homodyne detection allows to decrease $F_\mathrm{cond}$ compared to the value obtained for photon counting by choosing an appropriate homodyne angle $\varphi$. 
By this means, a nonclassical mean Fano factor $F_\mathrm{cond} < 1$ can be achieved even in the sideband-unresolved regime. 

Figure~\ref{fig:3}(c) shows that an imperfect detection, $\eta < 1$, reduces the effect such that $F_\mathrm{cond}$ tends towards $F_\mathrm{ss}$. 
If all knowledge of the measurement is discarded, $\eta = 0$, the continuous measurement is effectively absent and the relation $F_\mathrm{ss} = F_\mathrm{cond}$ holds. 
However, even for $50\,\%$ detection efficiency, a continuous measurement still lowers the Fano factor by about $25\,\%$. 
In Fig.~\ref{fig:4} we show that a continuous measurement is able to reduce the Fano factor more effectively for a small optomechanical coupling strength $g_0$. 
If $g_0$ becomes too large, $F_\mathrm{cond}$ tends towards the steady-state value. 
Finally, Fig.~\ref{fig:5} indicates that the reduction of $F_\mathrm{cond}$ with respect to $F_\mathrm{ss}$ is present at all considered mechanical temperatures.
However, in order to observe a nonclassical Fano factor $F_\mathrm{cond} < 1$, a small thermal phonon occupation $n_\mathrm{ph} \lesssim 1$ is required for the value of the mechanical damping considered here. 
Therefore, cryogenic temperatures or a precooling of the mechanics, e.g., using optomechanical cooling \cite{nature.475.359,nature.478.89}, are necessary. 
We stress that $n_\mathrm{ph}$ refers to an effective bath occupation number of such a combined mechanical and optical bath.

These numerical results can be qualitatively understood as follows. 
The steady-state phonon-number distribution $p_\mathrm{ss}(n)$ is given by an average over mechanical states at different amplitudes, each of them having a lower uncertainty in amplitude and phase than the steady-state distribution. 
In a minimal model, we assume the phonon distribution of these mechanical states to be a zero-mean distribution $p_\mathrm{bare}(n)$, which is centered around the instantaneous phonon-number expectation value $\erw{n}(t)$ and which is narrow enough to fulfill $\sqrt{\erw{n^2}_\mathrm{bare}} \ll \erw{n}_\mathrm{ss}$.
A continuous measurement on the optical cavity provides information on the state of the mechanical subsystem and allows us to track the diffusive movement $\erw{n}(t)$ of the mechanical state in phase space, cf.\ Fig.~\ref{fig:2}(a).
The measurement results $\erw{n}(t)$ are distributed according to the distribution $p_\mathrm{fluc}(n)$, whose Fano factor fulfills $F_\mathrm{fluc} < \erw{n}_\mathrm{fluc}$.
Consequently, we make the following ansatz for the steady-state phonon distribution: 
\begin{align}
	p_\mathrm{ss}(n) = \int_0^\infty \d n' \, p_\mathrm{fluc}(n') p_\mathrm{bare}(n - n') \fullstop
	\label{eqn:Discussion:Pss}
\end{align}
The instantaneous Fano factor $F(t \vert n)$ conditioned on a measurement result $n = \erw{n}(t)$ is $F(t \vert n) = \erw{n^2}_\mathrm{bare}/n$ and its average value is
\begin{align}
	F_\mathrm{cond} = \erw{\frac{1}{n}}_\mathrm{fluc} \erw{n^2}_\mathrm{bare} \fullstop
\end{align}
Comparing this result to the Fano factor obtained from Eq.~\eqref{eqn:Discussion:Pss}, we find
\begin{align}
	F_\mathrm{ss} \approx F_\mathrm{cond} + F_\mathrm{fluc} \comma
\end{align}
where the corrections are negligible in the limit of large a phonon-number expectation value $\erw{n}_\mathrm{fluc}$. 
Thus, the steady-state Fano factor is the sum of the conditional Fano factor of the mechanical states, which is resolved by a continuous measurement, and the Fano factor of the fluctuations of $\erw{n}(t)$, which smear out this information in the case of unconditional dynamics. 
This analytical result is well confirmed by our numerics. 

A continuous measurement on an optomechanical limit cycle is experimentally feasible with current technology. 
Optical homodyne detection on optomechanical systems is routinely done in experiments \cite{EPJD.22.131,PhysRevLett.111.163602,PhysRevLett.114.223601}.
The same holds for the realization of optomechanical limit cycles \cite{nphys.5.909,PhysRevLett.104.083901,PhysRevLett.109.233906,PhysRevLett.111.213902,nature.520.522}. 
To detect a sub-Poissonian mechanical state, optomechanical state-reconstruction techniques applicable to both the sideband-resolved and the sideband-unresolved regime are required. 
A proposal for state-reconstruction in the sideband-unresolved regime has been published recently \cite{arxiv.1709.01135}. 
In the sideband-resolved regime, several schemes are established and could be adapted to this setup \cite{anndp.527.15}.
For example, for optomechanical state-transfer protocols \cite{nature.482.63} it could be beneficial to add an auxiliary readout cavity to the system. 
A photon-counting measurement scheme has recently been applied to characterize the properties of a phonon laser \cite{nature.520.522}. 
Hence, the effect discussed in this article could already be verified in the sideband-resolved regime and there is a theoretical proposal on how to proceed in the case of the sideband-unresolved regime.

\section{Conclusion}
\label{sec:Conclusion}
In this article we numerically analyzed how homodyne and photon counting measurements on the optical cavity output decrease the mean mechanical Fano factor $F_\mathrm{cond}$ of an optomechanical system below the steady-state value $F_\mathrm{ss}$. 
In the sideband-resolved regime at small mechanical damping, the steady-state limit cycle is already nonclassical, $F_\mathrm{ss} < 1$, such that the additional benefit of a continuous measurement is small. 
However, in the sideband-unresolved regime, the mean Fano factor $F_\mathrm{cond}$ is drastically reduced compared to $F_\mathrm{ss}$ and the system is found in a nonclassical mechanical state for a macroscopic fraction of the observation time. 
In particular, for typical experimental parameters we observe a large decrease of the mechanical Fano factor. 
The effect is robust against imperfect detection, but a low effective thermal mechanical phonon number $n_\mathrm{ph}$ is necessary to observe nonclassical states.

We conclude that optical continuous measurements are a promising way to reduce amplitude fluctuations of the mechanical subsystem not only in the limit of cooling \cite{PhysRevLett.80.688,PhysRevA.60.2700,PhysRevB.68.235328,Nature.524.325}, but also for optomechanical limit cycles. 
This opens a route to the creation of nonclassical mechanical states in a new parameter range, namely, outside the sideband-resolved regime. 

\begin{acknowledgments}
We would like to thank M.\ Grimm for discussions.
Parts of the numerical calculations were performed at sciCORE scientific computing core facility at University of Basel \cite{scicore}.
This work was financially supported by the Swiss National Science Foundation (SNF) and the NCCR Quantum Science and Technology.
\end{acknowledgments}

\end{document}